\documentclass[aps,prl,twocolumn,superscriptaddress]{revtex4-1}

\usepackage{graphicx,bm,amssymb,multirow,amsmath,xcolor,pifont,soul}

\begin{document}

\title{Towards predictive many-body calculations of phonon-limited carrier mobilities in semiconductors}

\author{Samuel Ponc\'e}
\affiliation{Department of Materials, University of Oxford, Parks Road, Oxford, OX1 3PH, UK}
\author{Elena R. Margine}
\affiliation{Department of Physics, Binghamton University-SUNY, Binghamton, New York 13902, USA}
\author{Feliciano Giustino}
\email{feliciano.giustino@materials.ox.ac.uk}
\affiliation{Department of Materials, University of Oxford, Parks Road, Oxford, OX1 3PH, UK}

\newcommand{\mobun}{{cm$^2$/Vs}} 
\def\a{{\alpha}}
\def\b{{\beta}}
\def\ve{{\varepsilon}}
\def\w{\omega}
\def\bk{{\bf k}}
\def\bq{{\bf q}}
\def\bv{{\bf v}}
\def\bE{{\bf E}}
\def\bJ{{\bf J}}
\def\d{\delta}
\def\>{\rangle}
\def\<{\langle}
\def\D{\partial}
\def\kt{k_{\rm B}T}
\def\hbp{{\hat{\bf p}}}

\date{\today}

\begin{abstract}
We probe the accuracy limit of {\it ab initio} calculations of carrier mobilities in semiconductors,
within the framework of the Boltzmann transport equation. By focusing on the paradigmatic case of silicon,
we show that fully predictive calculations of electron and hole mobilities require many-body 
quasiparticle corrections to band structures and electron-phonon matrix elements, the inclusion 
of spin-orbit coupling, and an extremely fine sampling of inelastic scattering processes in momentum space.
By considering all these factors we obtain excellent agreement with experiment, and we identify the 
band effective masses as the most critical parameters to achieve predictive accuracy. Our findings 
set a blueprint for future calculations of carrier mobilities, and pave the way to engineering transport 
properties in semiconductors by design.
\end{abstract}

\maketitle

% word removed: 1+5+9+9+3+2+3

During the last decade, materials design guided by first-principles calculations has emerged 
as a powerful research strategy. Nowadays it is often possible 
to accurately predict ground-state properties of new materials {\it in silico}.
This information can be used to screen for promising new materials~\cite{Curtarolo2013,Jain2016}.
At variance with ground-state properties, the prediction and screening of materials properties involving 
electronic excitations is still in its infancy. For example charge and heat transport coefficients are 
typically evaluated using a combination of {\it ab initio} and semi-empirical approaches~\cite{Wang2011,Chen2013b,Hautier2013,Verstraete2014,Krishnaswamy2017}.
The reasons for this lag are that the evaluation of transport 
coefficients is considerably more challenging than total energies, the computational infrastructure is not 
yet fully developed, and the lack of a clear set of reference data for validation and verification~\cite{Lejaeghere2016}.

In this work, we focus on  
phonon-limited carrier mobilities in semiconductors. The theoretical framework for calculating mobilities is 
well established, and is rooted in the Boltzmann transport equation (BTE), as described in 
Refs.~\cite{Ziman1960,Grimvall1981, Mahan2000}. The BTE is a semiclassical, quasiparticle theory of 
electron transport, which can be rigorously derived from a many-body quantum-field theoretic framework by 
neglecting two-particle correlations~\cite{Kadanoff1962}. The key ingredients are the electronic band 
structures, the phonon dispersion relations, and the electron-phonon matrix elements. The calculations 
of these quantities have reached maturity~\cite{Giustino2017}, therefore 
there should be no fundamental obstacles towards predicting mobilities. However, already for the most 
studied semiconductor, silicon, one finds that (i) calculations of carrier 
mobilities are scarce, (ii) there is considerable scatter in the calculated data, and (iii) reproducing 
measured mobilities remains a challenge. For example, Refs.~\onlinecite{Zhou2015,Fiorentini2016,Li2015,Restrepo2009}
 calculate intrinsic electron mobilities at room temperature $\mu_{\rm e}=1550$, 
1750, 1860, and 1970~\mobun, respectively, while experiments are in the range 1300-1450~\mobun\ 
\cite{Canali1975,Norton1973,Sze2007}.

Motivated by these considerations, here we set to clarify the accuracy limit and the predictive
power of {\it ab initio} mobility calculations based on the BTE. We show that in order to correctly
reproduce experimental data we need to take into account GW quasiparticle corrections to the band structures
and the electron-phonon matrix elements, to include the spin-orbit splitting of the valence bands, and to properly converge the integrals over the Brillouin-zone. We also find that accurate band effective masses are absolutely critical to
reproduce measured mobilities. By considering all these aspects, we succeed in reproducing measured 
data with high accuracy, thus establishing unambiguously the predictive power of the {\it ab initio} BTE.

In a semiconductor the steady-state electric current $\bJ$ is related to the driving electric field
$\bE$ via the mobility tensors as: $J_\a = e \,(n_{\rm e} \,\mu_{{\rm e},\a\b} + n_{\rm h} 
\,\mu_{{\rm h},\a\b}) E_\b$, where Greek indices denote Cartesian coordinates.
In this expression $\mu_{{\rm e},\a\b}$, $n_{\rm e}$ and $\mu_{{\rm h},\a\b}$, $n_{\rm h}$ are the 
mobility and particle density of electrons and holes, respectively.
Within Boltzmann's transport formalism~\cite{Ziman1960} the current density is expressed as 
$J_\a = -e \,\Omega^{-1} \sum_n \Omega_{\rm BZ}^{-1}\int d\bk \, f_{n\bk} \,v_{n\bk,\a}$, where
$\Omega$ and $\Omega_{\rm BZ}$ are the volume of the crystalline
unit cell and the first Brillouin zone, respectively. The occupation factor $f_{n\bk}$ plays the role 
of a statistical distribution function, and reduces to the Fermi-Dirac distribution $f^0_{n\bk}$ in the 
absence of the electric field. The band velocity is given by $v_{n\bk,\a} = \hbar^{-1}\D \ve_{n\bk}/\D k_\a$, 
where $\ve_{n\bk}$ is the single-particle electron eigenvalue for the state $|n\bk\>$. 

Using these definitions, the electron mobility is obtained via the derivative of the current with respect
to the electric field:
$  \mu_{{\rm e},\a\b} = -\sum_{n\in {\rm CB}}\int \!d\bk\,\, v_{n\bk,\a}\D_{E_\b} f_{n\bk}\big/$ 
$\sum_{n\in {\rm CB}} \int \!d\bk\, f_{n\bk}^0$.
Here the summations are restricted to the conduction bands, and $\D_{E_\b}$ is short for $\D / \D E_\b$.
An analogous expression holds for holes. From this expression we see that in order to 
calculate mobilities we need to evaluate $\D_{E_\b} f_{n\bk}$, that is the linear response of the
distribution function $f_{n\bk}$ to the electric field $\bE$. This quantity can be computed starting from 
the BTE~\cite{Ziman1960}:
  \begin{eqnarray}
  && (-e) \bE \cdot \frac{1}{\hbar} \frac{\D f_{n\bk}}{\D {\bk}}  = \frac{2\pi}{\hbar}  
%  &&\frac{\D f_{n\bk}^0}{\D\ve_{n\bk}} \,\bv_{n\bk} \cdot (-e) \bE  = \frac{2\pi}{\hbar}
  \sum_{m\nu}\!\int\!\!\frac{d\bq}{\Omega_{\rm BZ}} \, |g_{mn\nu}(\bk,\bq)|^2  \nonumber \\  
  &&\,\,\,\, \times \big\{(1-f_{n\bk})f_{m\bk+\bq}   \d(\ve_{n\bk}-\ve_{m\bk+\bq} + \hbar\w_{\bq\nu})(1+ n_{\bq\nu})  \nonumber \\  
  &&\,\,\,\, \,\,\,\,         +(1-f_{n\bk})f_{m\bk+\bq}   \d(\ve_{n\bk}-\ve_{m\bk+\bq} - \hbar\w_{\bq\nu})n_{\bq\nu}  \nonumber  \\ 
  &&\,\,\,\, \,\,\,\,       -f_{n\bk}(1-f_{m\bk+\bq})   \d(\ve_{n\bk}-\ve_{m\bk+\bq} - \hbar\w_{\bq\nu})(1+ n_{\bq\nu})  \nonumber  \\  
  &&\,\,\,\, \,\,\,\,       -f_{n\bk}(1-f_{m\bk+\bq})   \d(\ve_{n\bk}-\ve_{m\bk+\bq} + \hbar\w_{\bq\nu})n_{\bq\nu} \big\}. 
   \label{eq.1}
  \end{eqnarray}
The left-hand side of Eq.~\eqref{eq.1} represents the collisionless term of Boltzmann's equation for 
a uniform and constant electric field, in the absence of temperature gradients and magnetic fields; 
the right-hand side represents the modification of the distribution function arising from electron-phonon 
scattering in and out of the state $|n\bk\>$, via emission or absorption of phonons with 
frequency $\w_{\bq\nu}$, wavevector $\bq$, and branch index $\nu$~\cite{Grimvall1981}. $n_{\bq\nu}$ is the Bose-Einstein distribution function.  
The matrix elements $g_{mn\nu}(\bk,\bq)$ in Eq.~(\ref{eq.1}) are the probability amplitude for scattering 
from an initial electronic state $|n\bk\>$ into a final state $|m\bk+\bq\>$ via a phonon $|\bq\nu\>$, 
as obtained from density-functional perturbation theory~\cite{Baroni2001,Giustino2017}. 
By taking derivatives of Eq.~\eqref{eq.1} with respect to $\bE$ we obtain an explicit
expression for the variation $\D_{E_\b} f_{n\bk}$:
  \begin{eqnarray}\label{eq.2}
  &&  \D_{E_\b}\!f_{n\bk} = \! e \frac{\partial f_{n\bk}^0}{\partial \varepsilon_{n\mathbf{k}}} v_{n\bk,\b} \tau_{n\bk}^0
\!+ \!\frac{2\pi\tau_{n\bk}^0}{\hbar}\!
     \sum_{m\nu}\!\!\int\!\!\!\frac{d\bq}{\Omega_{\rm BZ}} |g_{mn\nu}(\bk,\bq)|^2  \nonumber \\ 
   &&\quad   \times \left[ (1+n_{\bq\nu} - f_{n\bk}^0) \d(\ve_{n\bk}-\ve_{m\bk+\bq} + \hbar\w_{\bq\nu})  \nonumber \right. \\   
   &&\left. \quad  + (n_{\bq\nu} + f_{n\bk}^0)  \d(\ve_{n\bk}-\ve_{m\bk+\bq} - \hbar\w_{\bq\nu})\right]\,\D_{E_\b}\!f_{m\bk+\bq}, 
  \end{eqnarray}
having defined the relaxation time: 
  \begin{eqnarray}\label{eq.3}
  \frac{1}{\tau_{n\bk}^{0}} &=& \frac{2\pi}{\hbar} \sum_{m\nu}\!\int\!\!\frac{d\bq}{\Omega_{\rm BZ}} \, |g_{mn\nu}(\bk,\bq)|^2  \nonumber \\
  &\times& \left[(1-f_{m\bk+\bq}^0 +n_{\bq\nu}) \d(\ve_{n\bk}-\ve_{m\bk+\bq} - \hbar\w_{\bq\nu}) \right. 
   \nonumber \\
  &+& \left. (f_{m\bk+\bq}^0+n_{\bq\nu}) \d(\ve_{n\bk}-\ve_{m\bk+\bq} + \hbar\w_{\bq\nu})\right].    
  \end{eqnarray}
Equation~\eqref{eq.2} is the linearized BTE and is valid under the assumption that the energy
gained by a carrier accelerated by the electric field over the mean free path is much smaller
than the thermal energy, $e E_\b v_{n\bk,\b} \tau_{n\bk}^0 \ll k_B T$;
this assumption is verified in most semiconductors under standard operating conditions. 
This equation needs to be solved 
self-consistently for $\D_{E_\b} f_{n\bk}$, and is also referred to as the iterative BTE (IBTE). A simpler 
approach consists in neglecting the integral on the r.h.s. of Eq.~\eqref{eq.2}. In this case we 
obtain the variation $\D_{E_\b} f_{n\bk}$ without solving iteratively. 
It can be shown that the relaxation time $\tau_{n\bk}^{0}$ is related to the imaginary part of the Fan-Migdal electron 
self-energy~\cite{Giustino2017} via $1/\tau_{n\bk}^{0} = 2 \,{\rm Im}\,
\Sigma_{n\bk}^{\rm FM}$. 
Based on this analogy, in the following we refer to the approximation of 
neglecting the integral in Eq.~(\ref{eq.2}) as the `self-energy relaxation time approximation' (SERTA). 
In this approximation the mobility takes the simple form: 
  \begin{equation}\label{eq.4}
  \mu_{{\rm e},\a\b} = \frac{-e}{n_{\rm e}\,\Omega}\!\sum_{n\in {\rm CB}} 
  \int \!\frac{d\bk}{\Omega_{\rm BZ}}\,\, \frac{\partial f_{n\bk}^0}{\partial \varepsilon_{n\bk}} v_{n\bk,\a} \,v_{n\bk,\b}\,\tau_{n\bk}^{0}.
  \end{equation}
We perform calculations within density-functional theory (DFT), planewaves, and pseudopotentials 
using the \texttt{EPW} code~\cite{Ponce2016a} of the \texttt{Quantum ESPRESSO}
distribution~\cite{Giannozzi2017}, in conjunction with the \texttt{wannier90} library~\cite{Mostofi2014}. 
This approach employs a generalized Wannier-Fourier interpolation technique~\cite{Giustino2007} 
in order to obtain electron eigenvalues, phonon eigenfrequencies, and electron-phonon
matrix elements on dense Brillouin zone grids by means of maximally localized Wannier functions~\cite{Marzari2012}.
A fine sampling of the Brillouin zone is required because, at finite temperature, 
the Fermi level lies within the band gap, therefore we need to sample scattering processes 
taking place in the tails of the Fermi-Dirac distribution. 
%This can be seen from Eq.~\eqref{eq.4},
%wherein the Brillouin zone integral is weighted by $\partial f_{n\bk}^0/\partial \varepsilon_{n\bk}
% = -f_{n\bk}^0(1-f_{n\bk}^0)/k_{\rm B}T$. 
In our calculations the Fermi level is determined
in such a way that the net charge density at a given temperature, $n_{\rm e}-n_{\rm h}$, equals the 
doping level ($n_{\rm e}=n_{\rm h}$ for an intrinsic material). We now analyze in turn the key ingredients 
when calculating mobilities. We consider the paradigmatic case of silicon,
for which extensive experimental data are available.

\subsection{Brillouin-zone sampling} 
We find that in order to obtain reliable intrinsic mobilities it is necessary to employ 
extremely fine quasi-random grids, with a densified sampling around the band extrema. 
Convergence of mobility values to within 0.5\% 
is reached when using grids with 85K inequivalent $\bk$-points and 200K inequivalent $\bq$-points 
[white dot in Fig.~\ref{fig4}(a)]. Subsequent calculations in this article are performed using
these grids.
In Appendix Fig.~\ref{fig4}(b) we compare calculations of the intrinsic mobility of silicon within the SERTA and the IBTE
approaches. We find that the iterative solution of Eq.~\eqref{eq.2} leads to converged values which
are 6\% higher than the SERTA result for electrons, and 1\% lower for holes. Since the IBTE is drastically
more expensive because it requires homogeneous and commensurate grids~\cite{Li2015,Fiorentini2016}, 
in the following discussion we focus on SERTA calculations.
We use a finite broadening of 5~meV to evaluate the Dirac delta function in Eq.~\eqref{eq.3}.
The sensitivity of the results to the broadening parameter is analyzed 
in Fig.~\ref{fig5} of the Appendix.

\begin{figure}
  \centering
  \includegraphics[width=\columnwidth]{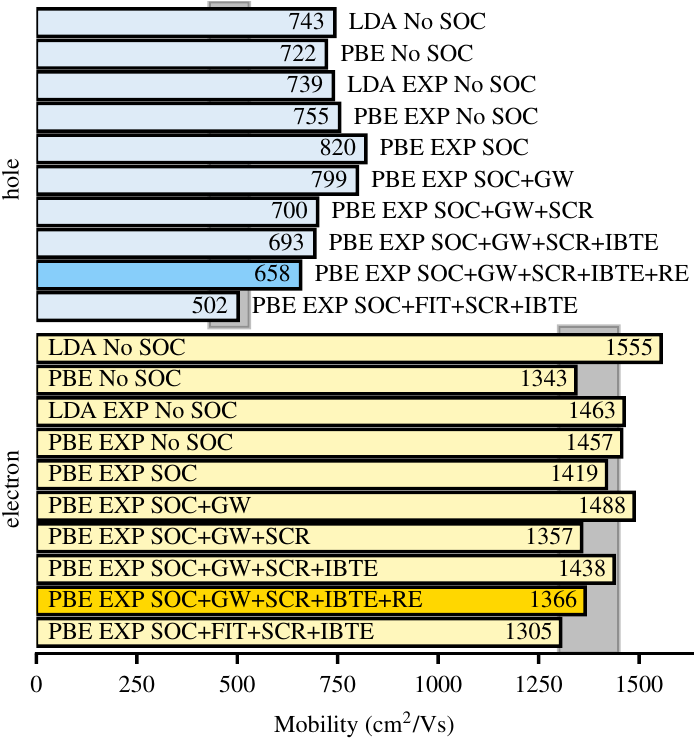}
  \caption{\label{fig1}
  Intrinsic electron and hole mobilities of silicon at 300~K, calculated using various levels of theory.
  The complexity of the theory increases as we move down the sequence of bars. The range of measured
  mobilities is indicated in light grey vertical bars. Our most accurate theoretical predictions are 
  $\mu_{\rm e} = 1366$~\mobun\ and $\mu_{\rm h} = 658$~\mobun;   by replacing the GW
  hole effective mass with the experimental value we obtain $\mu_{\rm e} = 502$~\mobun, in much better agreement
  with experiment. Key: SOC, spin-orbit coupling; EXP, experimental lattice parameter; 
  GW, calculations including quasiparticle corrections;
  SCR, electron-phonon coupling with corrected screening; 
  IBTE, iterative Boltzmann transport equation;
  RE, change of effective mass due to electron-phonon renormalization;
  FIT, band structures calculated from the measured effective masses.
  }
\end{figure}
\subsection{Exchange and correlation}
In order to investigate the effect of the DFT exchange and correlation we perform calculations
within both the local density approximation (LDA)~\cite{Ceperley1980,Perdew1981} and the 
generalized gradient approximation of Perdew, Burke, and Ernzerhof (PBE)~\cite{Perdew1996}, using scalar-relativistic pseudopotentials~\cite{Hamann2013}. Figure~\ref{fig1} shows that the intrinsic mobilities
at 300~K differ by 16\% between LDA and PBE for electrons, and by 3\% for holes. Closer inspection
shows that these differences arise primarily from the optimized lattice parameters obtained with
these functionals ($a=5.40$~\AA\ in LDA and 5.47~\AA\ in PBE). In fact, when using the experimental 
lattice parameter ($a=5.43$~\AA) the deviation between LDA and PBE mobilities reduces to 0.4\% for 
electrons and 2\% for holes (Fig.~\ref{fig1}). These results indicate that the choice of exchange and correlation
is not critical so long as accurate lattice parameters are employed.

\begin{figure}
  \centering
  \includegraphics[width=\columnwidth]{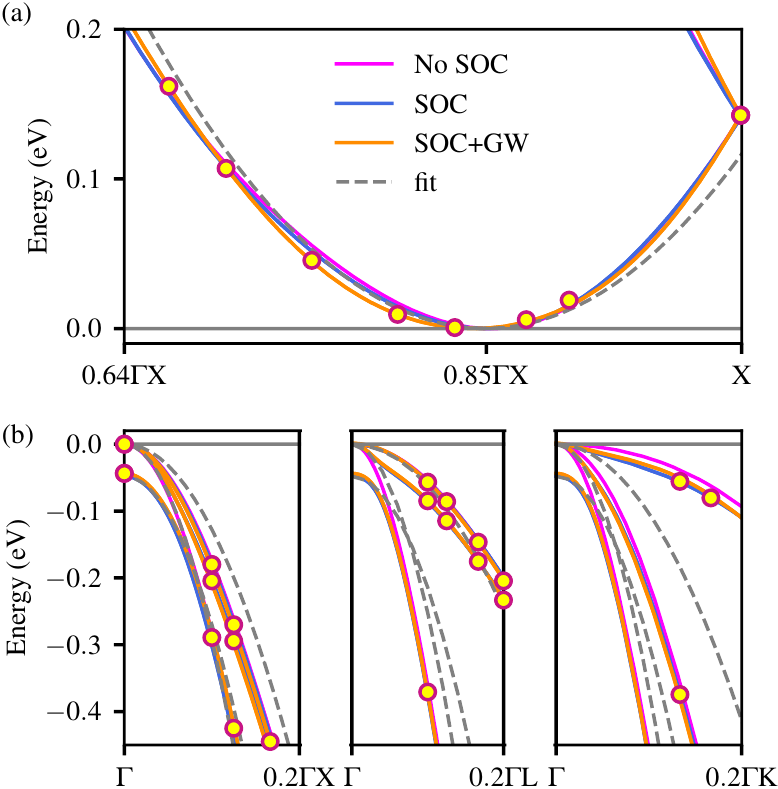}
  \caption{\label{fig2}
  (a) Conduction bands of silicon calculated within scalar-relativistic PBE (grey), fully-relativistic PBE
  (blue), the GW method (orange) and parabolic fit with measured effective masses (dashed).
  The zero of the energy axis is set to the conduction band minimum
  for clarity. (b) Valence bands of silicon, calculated within the same approximations as for (a), and
  shown using the same color code. The zero of the energy axis is set to the valence band top.
  In all panels the dots indicate explicit GW calculations carried out using uniform grids containing
  12$\times$12$\times$12 to 20$\times$20$\times$20 points. The GW bands in orange are obtained via
  Wannier interpolation. 
  }
\end{figure}

\subsection{Spin-orbit coupling}
Spin-orbit interactions in silicon are very weak~\cite{Yu2010}, therefore relativistic 
effects are usually neglected. However, here we find that
spin-orbit coupling is important for predictive calculations, yielding hole mobilities 9\% higher 
than non-relativistic calculations (Fig.~\ref{fig1}). This effect can be understood by considering 
the band structures in Fig.~\ref{fig2}(b). The spin-orbit interaction splits the six-fold 
degenerate states at the top of the valence bands, leading to the formation of two doubly-degenerate 
light-hole and heavy-hole bands, and one doubly-degenerate split-off hole band. As a result the effective 
mass of the light hole decreases (see Appendix Table~\ref{table1}), leading to a higher
mobility. On the other hand, Fig.~\ref{fig2}(a) shows that the conduction band bottom is 
relatively unaffected by spin-orbit coupling, and correspondingly the effect on the electron mobility is less pronounced 
(2.7\%).

\subsection{Many-body quasiparticle corrections}
Given the sensitivity of the calculated mobilities to the band extrema, we investigate the effect of
many-body correlations within the GW quasiparticle approximation. To obtain quasiparticle energies 
we use the \texttt{Yambo} code~\cite{Marini2009}; the values calculated
on a 12$\times$12$\times$12 uniform grid are then interpolated using the \texttt{EPW} code. Figure~\ref{fig2} shows the 
modification to the band extrema resulting from quasiparticle corrections. In the case of the valence
bands, quasiparticle corrections increase the mass of the light holes (see Table~\ref{table1} in Appendix); as a result the hole mobility decreases by 3\%, as shown in Fig.~\ref{fig1}. The opposite effect is observed for the
conduction bands where the electron mobility is increased by 5\%.

\begin{figure}
  \centering
  \includegraphics[width=\columnwidth]{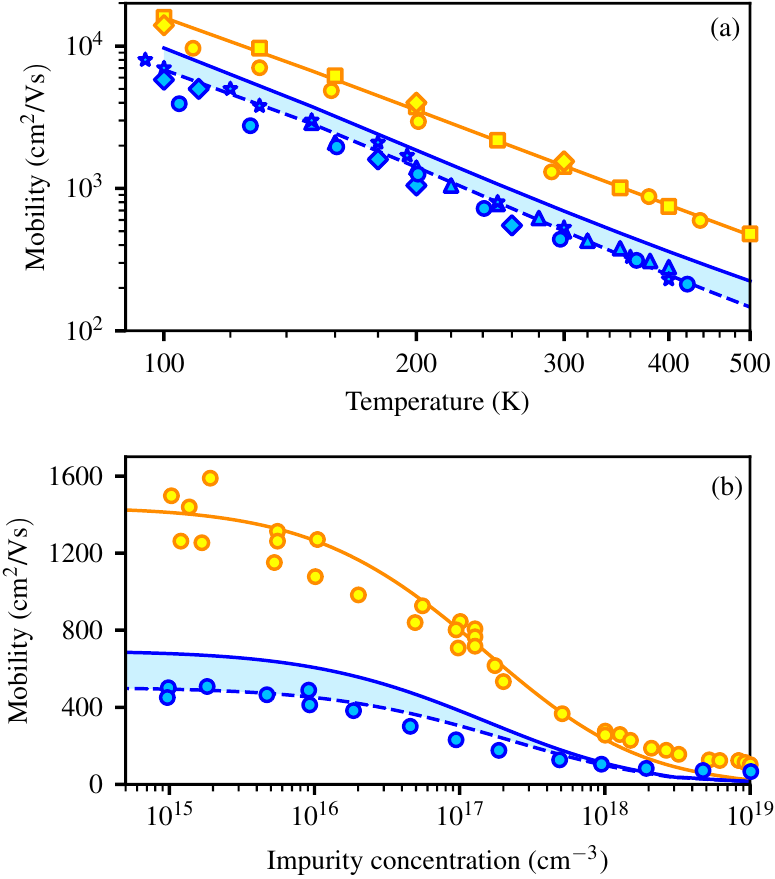}
  \caption{\label{fig3}
  (a) Comparison between calculated and measured intrinsic (low carrier concentration $\leq 10^{15}$ cm$^{-3}$) electron and hole mobilities of silicon,
  as a function of temperature. The calculations are performed using our best computational setup.
  The blue lines are for holes and the orange line is for electrons. In the case of holes we show
  both our best {\it ab initio} calculations (solid blue), and the results obtained by setting the
  hole effective mass to the experimental value (dashed blue). The shading is a guide to the eye.
  Experiments are from~\cite{Morin1954} ($\triangle$), \cite{Logan1960} ($\Diamond$),
  \cite{Ludwig1956} (\ding{73}), \cite{Jacoboni1977} ($\circ$), and \cite{Norton1973} ($\Box$).
  (b) Comparison between calculated and measured electron and hole mobilities of silicon at 300~K, 
  as a function of carrier concentration, using the same color code as in (a). 
  Experimental data are from~\cite{Jacoboni1977} ($\circ$).
  The impurity scattering is included via the model of Brooks and Herring with the Long-Norton correction~\cite{Brooks1951,Li1977} as described in the Appendix.
  }
\end{figure}

\subsection{Corrections to the DFT screening}
Another source of error in the DFT calculations of carrier mobilities is the overscreening of the
electron-phonon matrix elements $g_{mn\nu}(\mathbf{k,q})$ associated with the DFT band gap problem~\cite{Giustino2017}.
In fact, in the case of silicon DFT yields a static dielectric 
constant $\epsilon_{\rm DFT}^0 = 12.89$, which is higher than the measured value 
$\epsilon_{\rm exp}^0 = 11.94$~\cite{Sze2007}. In order to overcome this issue it is 
necessary to modify the screening in the calculation of phonon dispersion relations. Since this is 
computationally prohibitive, here we take a simpler approach and renormalize the matrix 
elements as follows: $g^\prime_{mn\nu}(\mathbf{k,q}) = g_{mn\nu}(\mathbf{k,q}) 
\left[\epsilon_{\rm DFT}(|\bq|)/ \epsilon_{\rm exp}(|\bq|)\right]$.
Here $\epsilon_{\rm exp}$ is meant to be the most accurate description of the screening that we can afford,
and we are neglecting local-field effects which should yield an error on the order of a few percent~\cite{Hybertsen1987}. 
For practical purposes we replace the 
dielectric functions by an analytic expression~\cite{Resta1977}, where the only input parameter is the head of
the dielectric matrix. The validity of this procedure
is demonstrated in Appendix Fig.~\ref{fig6} using explicit
calculations in the random-phase approximation. This correction to the matrix elements leads
to a decrease of the electron and hole mobilities by 8.8\% and
12.4\%, respectively, as shown in Fig.~\ref{fig1}.

\subsection{Thermal expansion and electron-phonon renormalization} We computed the effect of 
thermal lattice expansion on the DFT eigenenergies using the \texttt{thermo\_pw} code~\cite{thermopw} within the quasi-harmonic approximation 
and concluded that this effect is negligible, see Appendix Figs.~\ref{fig7}-\ref{fig8}.
We also determined the electron-phonon renormalization of the effective masses using data
from Ref.~\onlinecite{Ponce2015}. This effect increases the masses by $\sim$3\%, and results into
a decrease of the mobilities by $\sim$5\%. 

%electron:1360 \cite{Cronemeyer1957}, 1350\cite{Ludwig1956},1430\cite{Li1977}, 1450\cite{Jacoboni1977}
%hole:510\cite{Cronemeyer1957},  480 \cite{Ludwig1956}, 450\cite{Jacoboni1977},445\cite{Dorkel1981}
After considering all the effects discussed so far, and after accounting for the corrections to the SERTA
results arising from the solution of the complete IBTE, our most accurate theoretical mobilities
at 300~K are $\mu_{\rm e} = 1366$~\mobun\ and $\mu_{\rm h} = 658$~\mobun.  
These values are to be compared to the measured drift mobilities $\mu_{\rm e}^{\rm exp} = 1350$-1450~\mobun\
\cite{Ludwig1956,Cronemeyer1957,Li1977,Jacoboni1977} 
and $\mu_{\rm h}^{\rm exp} = 445$-510~\mobun\ \cite{Dorkel1981,Jacoboni1977,Ludwig1956,Cronemeyer1957} 
(Fig.~\ref{fig1}). From the comparison with experiment
we see that by pushing the theory to its limits we can obtain electron mobilities in very good agreement
with experiment. On the contrary, the hole mobility are still approximately 30\% above the
measured range. This discrepancy can be traced back to the underestimation of the [100] heavy hole
effective masses within the GW approximation. In fact, by repeating the calculation using the 
experimental hole effective mass instead of the GW mass, we obtain a hole mobility $\mu_{\rm h}^\prime 
= 502$~\mobun, this time in very good agreement with experiment as shown in Fig.~\ref{fig1}. 
This result leads us to conclude 
that the effective mass plays an absolutely critical role in mobility calculations. Our finding can be understood
by considering that the mobility varies with the effective mass as $\mu = (m^*)^{-p}$ with $p$ 
being a coefficient between 1 and 2.5~\cite{Bardeen1950,Blatt1957,Keyes1959}; as a result a 20\% error
 in the effective mass leads to an
error in the mobility of up to 60\%. This finding highlights the critical role of accurate
calculations of quasiparticle band structures, and raises the question on whether the standard GW
method and pseudopotential calculations (see Table~\ref{table2}) are 
sufficient for delivering predictive mobilities.

Using the best possible computational setup we can now compare our calculations with experiment
over a range of temperatures and doping levels. Figure~\ref{fig3}(a) shows the intrinsic electron
and hole mobilities of silicon between 100~K and 500~K. In the case of the hole mobilities
we show both our best {\it ab initio} results (solid line), as well as those re-calculated using 
the experimental effective masses (dashed line). Overall, the agreement between our calculations
and experiment is very good throughout the entire temperature range. Figure~\ref{fig3}(b) shows
a comparison between calculated and measured mobilities at 300~K, as a function of carrier concentration
between $10^{15}$ and $10^{19}$~cm$^{-3}$.
In this case, in addition to the \textit{ab initio} electron-phonon scattering, 
we used the semi-empirical model of Brooks and Herring
with the Long-Norton correction~\cite{Brooks1951,Li1977} to account for impurity scattering
(see Appendix for details). 
Also in this case we find very good agreement with experiment, although the contribution 
of impurity scattering is evaluated semi-empirically. 

In conclusion, we pushed the accuracy of transport calculations within the BTE formalism to its limits,
and we demonstrated that this approach can deliver predictive accuracy for a prototypical semiconductor. 
Our findings raise two important 
questions for future work on transport in semiconductors: (i) the present formalism yields results
which fall within the experimental uncertainty. In order to enable further progress in this area
it will be important to produce a high-quality experimental data from single-crystal
samples. (ii) An unexpected challenge that we faced is to perform accurate \textit{ab initio}
calculations of effective masses. Going forward it will be important to establish whether the GW method 
and pseudopotential calculations can provide
effective masses with the accuracy required for predictive mobility calculations. 
%In hindsight, it is surprising
%that such a basic property of semiconductors has received very little if no attention at all so far.
Meanwhile, the present work opens the way to predictive calculations of mobilities and lays the
groundwork for the {\it ab initio} design of semiconductor devices.

\paragraph{Note added.} After submission of this work, a related calculation for Si
was reported, where the authors found a significant increase in Si hole mobility with SOC and no effect
from SOC on the electron mobility in line with our results~\cite{Ma2018}.

\begin{acknowledgments}
We acknowledge fruitful discussions with C. Verdi, M. Schlipf and W. Li.
This work was supported by the Leverhulme Trust (Grant RL-2012-001), the UK EPSRC Research Council 
(grants No. EP/J009857/1 and EP/M020517/1), the EU H2020
programme under grant No.~696656 GrapheneCore1, the University of Oxford Advanced Research Computing (ARC) 
facility (http://dx.doi.org/810.5281/zenodo.22558), the ARCHER UK National Supercomputing Service under 
the AMSEC and CTOA projects, PRACE DECI-13 resource Cartesius at SURFsara, and the PRACE DECI-14 resource 
Abel at UiO. E.R.M. acknowledges the NSF support (Award No. OAC-1740263).
\end{acknowledgments}

\section{Appendix}

\subsection{Computational Methods}

In this work we use norm-conserving pseudopotentials with planewave kinetic energy cutoffs of 45~Ry 
and 35~Ry for LDA and PBE calculations, respectively. The phonon dispersion relations are evaluated
using density-functional perturbation theory~\cite{Baroni2001}, starting from a 
6$\times$6$\times$6 uniform grid of $\bq$-points. An 18$\times$18$\times$18 uniform grid of $\bk$-points 
was required to correctly obtain vanishing Born effective charges.
Representative phonon dispersion relations obtained
within this setup can be found in Ref.~\cite{Ponce2016a}. 
The coarse grids for the electron-phonon interpolation required 12$\times$12$\times$12 $\bk$-points and 
6$\times$6$\times$6 $\bq$-points. Such a dense $\bk$-grid was needed to obtain a good Wannier interpolation
of the conduction bands, since the minimum is along the $\Delta$ line (approx.~0.85$\Gamma$X) 
and does not fall on a high-symmetry point.

For the self-consistent, iterative solution of the Boltzmann transport equation (IBTE) we employ
uniform Brillouin-zone grids, and the $\bq$-point sums are restricted to the irreducible wedge
of the Brillouin zone using crystal symmetry operations. The IBTE is solved using homogeneous and commensurate
$\bk$- and $\bq$-point grids since the variations $\D_{E_\b} f_{n\bk}^{i+1}$ at the $(i+1)$-th
iteration require the knowledge of the variations $\D_{E_\b} f_{n\bk+\bq}^{i}$ at the $i$-th iteration,
see Eq.~\eqref{eq.2}.

For the direct solution of the BTE within the self-energy relaxation time approximation (SERTA)
the Brillouin zone grids do not need to be commensurate. In this case, in order to improve the sampling 
accuracy, we employ quasi-random Sobol sequences of $\bk$- and $\bq$-points. Following recommended
practice, we skip the first 1000 elements of a sequence and we retain one element every 100 of the
remainder~\cite{Bratley1988}; furthermore we employ a linear scramble and shift of the resulting
sequence, using standard routines from \texttt{Matlab~R2015a}~\cite{Hong2003}. As a further
refinement we replace the homogeneous Sobol weights using a Voronoi
triangulation with the code \texttt{Voro++}~\cite{Rycroft2009}. In the Voronoi triangulation we take
into account the periodicity of the Brillouin zone by building periodic replicas of the random grid
in neighboring reciprocal unit cells. For the $\bk$-point grid we also densify the distribution
around the band extrema, in order to capture the fine features of the scattering near the band edges.
This is achieved by generating additional random points with the Lorentz distribution
$1/(1+|\bk-\bk_0|^2/\gamma^2)$ and by recomputing the Voronoi weights of the resulting grid. Here $\bk_0$ indicates the location of the band extrema and $\gamma=0.008$~\AA$^{-1}$.

\begin{figure}
  \centering
  \includegraphics[width=\columnwidth]{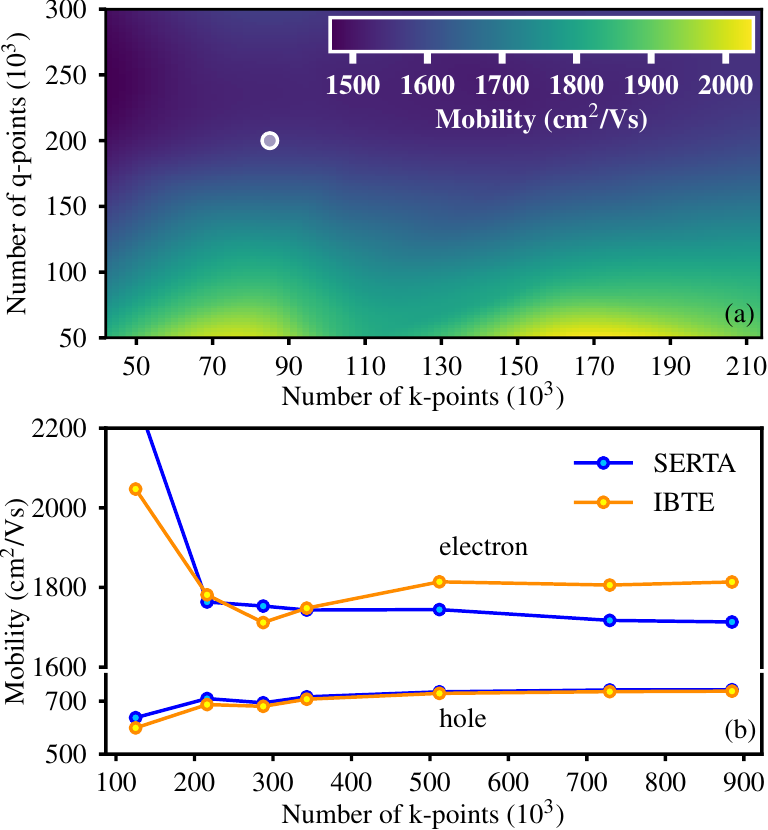}  
  \caption{\label{fig4}
  (a) Sensitivity of the intrinsic electron mobility of silicon at 300~K with respect to the sampling 
  of electron ($\bk$) and phonon ($\bq$) wavevectors in the Brillouin zone. The calculations are
  performed within the SERTA approximation, using a densified Lorentzian distribution of $\bk$-points around 
  the conduction band minima, and a Sobol quasi-random sampling of the $\bq$-points. The white dot
  indicates the setup used in the calculations reported in this article.
  (b) Comparison between the rate of convergence of the intrinsic electron and hole mobilities of silicon 
  using the SERTA and the IBTE approaches, at 300~K. In this case we use uniform grids, with the $\bk$-point
  mesh being twice as dense as the $\bq$-point mesh in each direction.
  }
\end{figure}

Figure~\ref{fig4}(a) shows the convergence of the intrinsic mobility of silicon at 300~K with respect to
the number of electron and phonon wavevectors in the Brillouin zone within the SERTA approximation. 
Figure~\ref{fig4}(b) shows the comparison between calculations of the intrinsic mobility of silicon within the SERTA and the IBTE approaches.

The GW calculations are performed starting from the PBE band structure and using the experimental
lattice parameters on a 12$\times$12$\times$12 $\bk$-point grid. To obtain direct and indirect band gaps converged to within 5~meV
we use 120~bands and a planwaves cutoff of 15~Ry for the dielectric matrix. The renormalization
of the band velocity is evaluated as in Ref.~\cite{Rohlfing2000}: 
$\< \psi_{n\bk} | \hbp | \psi_{m\bk}\>_{\rm GW} = 
[(\ve_{n\bk}^{\rm GW}-\ve_{m\bk}^{\rm GW})/(\ve_{n\bk}^{\rm DFT}-\ve_{m\bk}^{\rm DFT})]
\< \psi_{n\bk} | \hbp | \psi_{m\bk}\>_{\rm DFT}$, where $\hbp$ indicates the momentum operator.
When $n=m$ the previous expression is replaced by 
$\< \psi_{n\bk} | \hbp | \psi_{n\bk}\>_{\rm GW} = \< \psi_{n\bk} | \hbp | \psi_{n\bk}\>_{\rm DFT}$.

For completeness the effective masses computed within scalar-relativistic DFT, fully-relativistic DFT,
and including GW quasiparticle corrections are reported in Table~\ref{table1}. 
We also show in Table~\ref{table2} the effective masses calculated without SOC at the experimental lattice parameter with 
two different types of pseudization (norm-conserving and ultrasoft), and two exchange and correlation 
functionals (LDA and PBE).

In order to calculate mobilities using band structures as close as possible to experiments
(i.e. the lowermost bars in Fig.~\ref{fig1}),
we repeated the calculations using the low-energy dispersion relations parametrized 
in Refs.~\onlinecite{Dresselhaus1955,Yu2010} starting from the measured effective masses:
\begin{align}
\varepsilon_{\text{cb}} =& \frac{\hbar^{2}(k_x-k_{0,x})^2}{2m_{||}}+ \frac{\hbar^{2}(k_y-k_{0,y})^2}{2m_{\perp}} \nonumber\\
                        &  + \frac{\hbar^{2}(k_z-k_{0,z})^2}{2m_{\perp}} + \varepsilon_{\rm c}, \\
\varepsilon_{\text{hh}} =& Ak^2 + [B^2k^4 + C^2(k_x^2k_y^2 + k_y^2k_z^2 + k_z^2k_x^2)]^{1/2},\\
\varepsilon_{\text{lh}} =& Ak^2 - [B^2k^4 + C^2(k_x^2k_y^2 + k_y^2k_z^2 + k_z^2k_x^2)]^{1/2},\\
\varepsilon_{\text{so}} =& -\frac{k^2\hbar^2}{2m_{\text{so}}} - \varepsilon_{\text{so}},
\end{align}
where $m_{||}=0.98m_0$ ($m_0$ is the free electron mass), 
$m_{\perp}=0.19m_0$, $m_{\text{so}}=0.23m_0$,  $\bk_0$ denotes the wavevectors of 
the conduction band minima, and $\varepsilon_{\rm c}$ is the conduction band bottom. 
The coefficients are $A=-4.1\,\hbar^2/2m_0$, $B=-1.6\,\hbar^2/2m_0$ 
and $C=3.3\,\hbar^2/2m_0$~\cite{Dresselhaus1955,Yu2010} and 
$\varepsilon_{\text{so}}$ = 48~meV.

\begin{table}
  \begin{tabular}{c c c c c c }
  \toprule
  Band & Direction & \multicolumn{3}{c}{Present calculations} & Expt. \\
  \multicolumn{2}{c}{}   & No SOC & SOC & SOC+GW &     \\
  \hline\\[-8pt]
  \multirow{3}{*}{Split-off hole}
  & [100]                & 0.167  & 0.224 &  0.226 &  0.23  \\
  & [111]                & 0.094  & 0.227 &  0.227 &  0.23  \\
  & [110]                & 0.106  & 0.227 &  0.225 &  0.23  \\
  \multirow{3}{*}{Light hole}
  & [100]                & 0.253  & 0.189 &  0.202 &  0.17 \\
  & [111]                & 0.682  & 0.131 &  0.132 &  0.16 \\
  & [110]                & 0.266  & 0.140 &  0.140 &  0.16 \\
  \multirow{3}{*}{Heavy hole}
  & [100]                & 0.271  & 0.256 &  0.243 &  0.46 \\
  & [111]                & 0.694  & 0.654 &  0.643 &  0.56 \\
  & [110]                & 2.868  & 0.521 &  0.512 &  0.53 \\
  \multirow{2}{*}{Electron}  
          & long.  & 0.798  & 0.824 &  1.090  & 0.98 \\ 
          & trans. & 0.188  & 0.190 &  0.186  & 0.19 \\
  \botrule 
  \end{tabular}
  \caption{\label{table1}
  Comparison between calculated and measured effective masses of silicon, in units of the electron mass.
  The experimental data are from Refs.~\onlinecite{Dexter1954,Sze2007,Yu2010}.
  }
\end{table}

\begin{table}
  \begin{tabular}{c c c c c c }
  \toprule
  Band & Direction       & LDA-US       & LDA          & PBE        &  Expt. \\
  \hline\\[-8pt]
  \multirow{3}{*}{Split-off hole}
  & [100]                & 0.170 & 0.168 & 0.167  &   0.23  \\
  & [111]                & 0.098 & 0.098 & 0.094  &   0.23  \\
  & [110]                & 0.111 & 0.109 & 0.106  &   0.23  \\
  \multirow{3}{*}{Light hole}
  & [100]                & 0.248 & 0.265 & 0.253  &   0.17 \\
  & [111]                & 0.551 & 0.655 & 0.682  &   0.16 \\
  & [110]                & 0.271 & 0.278 & 0.266  &   0.16 \\
  \multirow{3}{*}{Heavy hole}
  & [100]                & 0.271 & 0.276 & 0.271  &   0.46 \\
  & [111]                & 0.635 & 0.678 & 0.694  &   0.56 \\
  & [110]                & 2.158 & 2.170 & 2.868  &   0.53 \\
  \multirow{2}{*}{Electron}  
          & long.  & 0.755 & 0.735 & 0.771  &   0.98 \\ 
          & trans. & 0.182 & 0.185 & 0.188  &   0.19 \\
  \botrule 
  \end{tabular}
  \caption{\label{table2}
  Comparison between effective masses calculated using different types of pseudopotentials and exchange-correlation functionals without SOC, in units of the electron mass.
  The experimental data are from Refs.~\onlinecite{Dexter1954,Sze2007,Yu2010}.
  }
\end{table}

%%%%%%%%%%%%%%%%%%%%%%%%%%%%%%%%

\subsection{Broadening of Dirac delta functions}
The numerical evaluation of phonon-limited mobilities using Eqs.~\eqref{eq.2}-\eqref{eq.4} requires one to
replace the Dirac delta functions in Eqs.~\eqref{eq.2}-\eqref{eq.3} by Lorentzian functions with finite broadening $\eta$: 
$\pi\,\d(\ve_{n\bk}\pm\hbar\w_{\bq\nu}-\ve_{m\bk+\bq}) \rightarrow {\rm Im}\,(\ve_{n\bk}\pm\hbar
\w_{\bq\nu}-\ve_{m\bk+\bq}-i\eta)^{-1}$. This procedure makes the calculated mobility dependent
on the broadening parameter, hence it is important to check how sensitive are the results to the 
choice of $\eta$.

\begin{figure}
  \centering
  \includegraphics[width=\columnwidth]{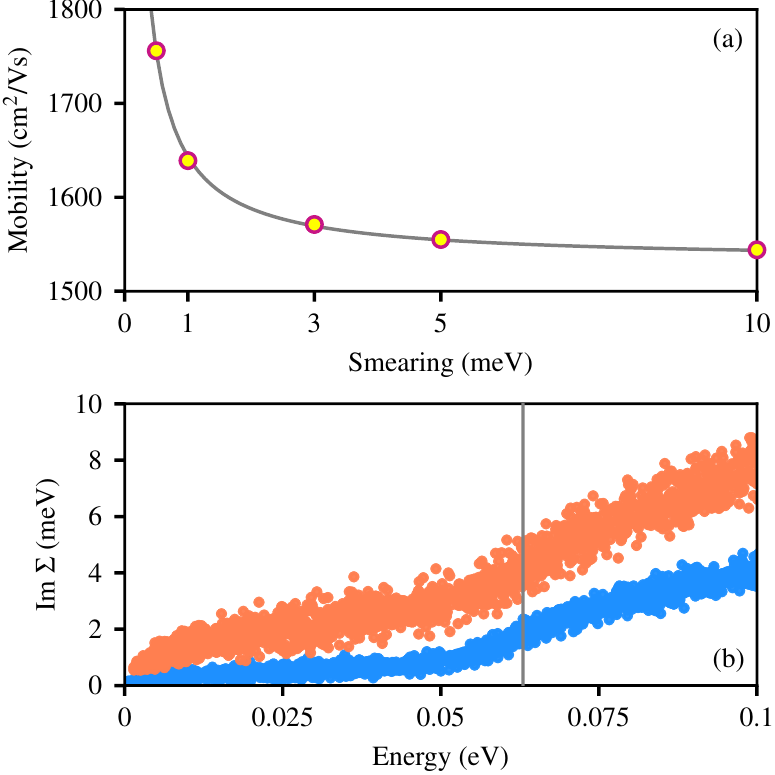}
  \caption{\label{fig5}
  (a) Intrinsic electron mobility of silicon at 0~K, 
  calculated as a function of the broadening parameter $\eta$
  (dots). The grey thin line is a guide to the eye and was obtained by fitting the data points using
  $\mu = {\rm const}/ \eta$.
  (b) Electron quasiparticle linewidths in silicon arising from the electron-phonon interaction,
  calculated at 0~K (blue dots) and 300~K (orange dots). The zero of the horizontal energy axis is set
  to the conduction band minimum. The vertical grey line indicates the energy of the highest optical
  phonon in silicon.
  }
\end{figure}

Figure~\ref{fig5}(a) shows the intrinsic electron mobility of silicon at 0~K, evaluated as a function of $\eta$.
From this figure we see that the mobility tends to diverge towards $+\infty$ as $\eta\rightarrow 0$.
This trend can be rationalized by noting that the mobility is directly proportional to the relaxation
time [cf.\ Eq.~\eqref{eq.4}], and the relaxation time due to acoustic phonon scattering in a
non-polar semiconductor is inversely proportional to the temperature~\cite{Bardeen1950}. As a result, we expect that the phonon-limited
mobility will increase indefinitely as $\eta$ becomes smaller and the Lorentzian approaches the Dirac 
delta function. This observation is in agreement with the explicit calculations in Fig.~\ref{fig5}(a).

This behavior poses a problem when one has to decide which broadening parameter to use in the
calculations. As a general rule here we set $\eta$ to the smallest possible value where the
curve $\mu$ vs.\ $\eta$ is relatively flat, so that our results are insensitive to this choice.
Based on Fig.~\ref{fig5}(a), we use $\eta = 5$~meV in all calculations presented in the article. This choice
is consistent with the notion that real quasiparticles do not have an infinite lifetime as it is assumed 
in the BTE formalism, but have a finite lifetime due to electron-electron and electron-phonon interactions.
In Fig.~\ref{fig5}(b) we show our calculated quasiparticle broadening from electron-phonon interactions at
0~K and 300~K. It can be seen that at 300~K the broadening reaches values up to 4-5~meV 
for quasiparticle energies located one phonon energy away from the band bottom (the highest phonon
energy in silicon is $\sim$63~meV). These values are consistent with our choice of broadening parameter.

\subsection{Screening of the electron-phonon matrix elements}

The strategy that we used to correct for the DFT overscreening of the electron-phonon matrix elements
consists of un-screening the matrix elements via the DFT dielectric function, so as to obtain
the bare matrix elements, and then screening the bare matrix elements using the best possible
dielectric function. The dielectric function can be factored out of the integral in the matrix element
if we neglect local field effects.
Since local field effects are known to decrease the head of the dielectric function of silicon by 10\%
and the body of the dielectric function 
is typically one or two orders of magnitude smaller than the head~\cite{Hybertsen1987}, we expect 
to make an error on the order of a few percent. 

In order to perform this operation for a large number of phonon wavevectors
we use the Thomas-Fermi model dielectric function of Ref.~\onlinecite{Resta1977}:
  \begin{equation}\label{eq.model}
  \epsilon(q) = \frac{k_0^2+q^2}{k_0^2 \sin (qR)/(qR\,\epsilon_0) + q^2 },
  \end{equation}
where $q=|\bq|$, $\epsilon_0$ is the macroscopic (electronic) dielectric constant. $k_0$ are $R$ are obtained from the valence electron density $\rho$ as 
$k_0^2 = 4(3\pi^2 \, \rho)^{1/3}/\pi$ and $\sinh (k_0R)/k_0R = \epsilon_0$. The only free parameter of the model is $\epsilon_0$. 

\begin{figure}
  \centering
  \includegraphics[width=\columnwidth]{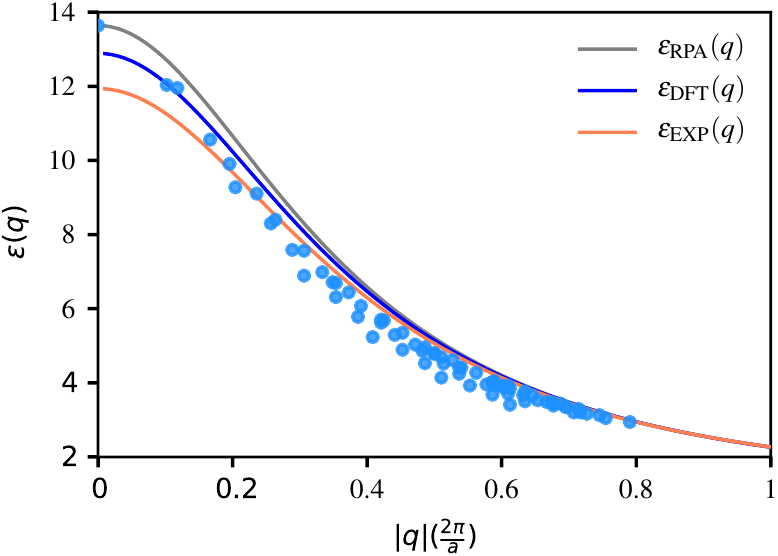}
  \caption{\label{fig6}
  Comparison between the diagonal part of the RPA dielectric matrix of silicon (blue dots) and the Thomas-Fermi
  model of Ref.~\onlinecite{Resta1977} (grey line). 
  We also show the model dielectric function using the DFT dielectric constant $\varepsilon(0)=12.89$ (blue line) and 
  experimental dielectric constant $\varepsilon(0)=11.94$ (orange line).
  }
\end{figure}

We test the validity of this model by computing the dielectric matrix within the random phase approximation (RPA),
using a 12$\times$12$\times$12 unshifted grid (corresponding to 72~inequivalent wavevectors).
Figure~\ref{fig6} shows a comparison between the model dielectric function of Eq.~\eqref{eq.model} 
and the RPA calculation,
after matching $\epsilon_0$ to the head of the RPA dielectric matrix. We see that Eq.~\eqref{eq.model} reproduces well the RPA
screening, therefore it is sensible to use it in the renormalization of the electron-phonon
matrix elements. In this work we renormalized the matrix elements by setting $\epsilon_0$ to the
experimental dielectric constant of silicon (11.94). 

\subsection{Brooks-Herring model for impurity scattering}

In order to account for impurity scattering in Fig.~\ref{fig3}(b), we use the
semi-empirical model developed by Brooks and Herring~\cite{Brooks1951,Li1977}.
In this model the mobility $\mu_{\rm i}$ is evaluated analytically by taking into
account quantum-mechanical scattering rates, spherical energy surfaces, negligible 
electron-electron interactions, and complete ionization of the impurities. 
The explicit expression of the hole mobility is:
  \begin{equation}\label{BHequation}
  \mu_{\rm i} = \frac{2^{7/2} \epsilon_s^2 (\kt)^{3/2} }{\pi^{3/2} e^3 \sqrt{m_d^*} \,n_{\rm i} 
  G(b)} \quad \bigg[\frac{\rm{cm^2}}{\rm{Vs}}\bigg],
  \end{equation}
where $G(b) = \ln(b+1) - b/(b+1)$, $b = 24\pi m_d^* \epsilon_s (\kt)^{2}/e^2 h^2 n'$, and
$n'=n_{\rm h}(2-n_{\rm h}/n_{\rm i})$.
Here $m_d^*=0.55 m_0$ is the density-of-state effective mass for the holes~\cite{Balkanski2000}, $n_{\rm h}$ and $n_{\rm i}$ are the hole densities and 
the density of ionized impurities [impurity concentration in Fig.~\ref{fig3}(b)], 
respectively, $\epsilon_s=11.9\epsilon_0$ is the dielectric constant, $\epsilon_0$ is the
permittivity of vacuum, and $h$ is Planck's constant. In the above expressions, 
the concentrations are expressed in cm$^{-3}$, and the temperature $T$ is in K.

In the case of silicon, Eq.~\eqref{BHequation} cannot be used for the electron mobility 
because the electron mass is highly anisotropic and leads to incorrect results. To account for the electron mass 
anisotropy, we instead used the Long-Norton mobility expression~\cite{Norton1973,Li1977}:
  \begin{equation}
  \mu_{\rm i}^{\rm{LN}} = \frac{7.3\cdot 10^{17} T^{3/2} }{n_i G(b)} \quad \bigg[\frac{\rm{cm^2}}{\rm{Vs}}\bigg],
  \end{equation}
where the electron density-of-state effective mass is $m_d^*=1.08 m_0$~\cite{Sze2007}.

 Finally, the mobility including phonon ($\mu_l$) and impurity ($\mu_i$) scattering can be computed using the 
 mixed-scattering formula~\cite{Li1977}:
  \begin{equation}
  \mu =\mu_l \Big[ 1 + X^2\{ \text{ci}(X)\cos(X) + \sin(X)(\text{si}(X)-\frac{\pi}{2})\}  \Big], 
  \end{equation}
where $X^2 = 6\mu_l/\mu_i$ and ci(X) and si(X) are the cosine and sine integrals.

\subsection{Effect of thermal lattice expansion}

Within the quasi-harmonic approximation~\cite{Baroni2010}, the Helmholtz free energy of a 
cubic crystal is given by~\cite{Palumbo2017}:
\begin{equation}
F(T,V) = U(V) + F^{\rm{vib}}(T,V) + F^{\rm{el}}(T,V),
\end{equation}
where $U$ is the static energy at 0~K, $F^{\rm{vib}}$ is the contribution due to lattice vibration 
and $F^{\rm{el}}$ the energy due to electronic thermal excitations. 
We rely on the adiabatic approximation to treat each term independently. 
The vibrational Helmholtz free energy per cell is given in the harmonic approximation by~\cite{Palumbo2017}: 
\begin{align}
F^{\rm{vib}}(T,V) =& \frac{1}{2N} \sum_{\mathbf{q},\nu} \hbar \omega_{\mathbf{q},\nu}(V) \nonumber \\
+& \frac{k_B T}{N}  \sum_{\mathbf{q},\nu} \ln \bigg[ 1 - \exp\Big( \frac{-\hbar \omega_{\mathbf{q},\nu}(V)}{k_B T} \Big) \bigg],
\end{align}
where $N$ is the number of $\bf q$-points, the first term is the contribution to the zero-point energy and the second term is the 
phonon contribution at finite temperature. $F^{\rm{el}}$ can be neglected as the band
gap is much larger than thermal energies.

\begin{figure}
  \centering
  \includegraphics[width=\columnwidth]{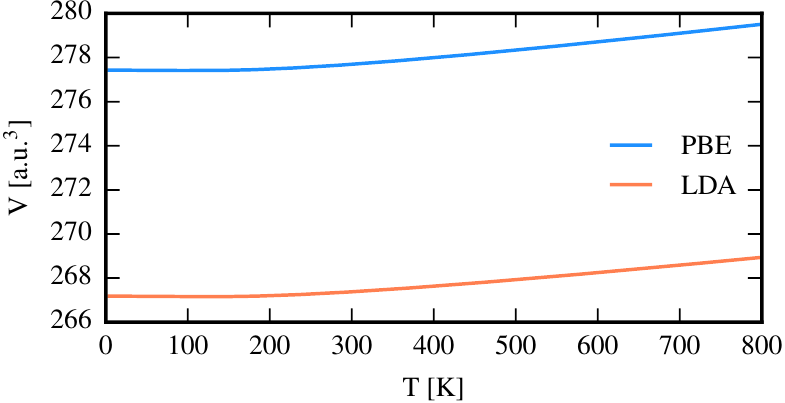}
  \caption{\label{fig7}
  Variation of volume with temperature due to thermal expansion using the LDA or PBE exchange correlation functionals.
  }
\end{figure}

The energy minimum of $U(V) + F^{\rm{vib}}(T,V)$ at a given temperature corresponds to zero pressure and gives the variation of volume
with temperature due to thermal expansion. 
To perform those calculations we used the \texttt{thermo\_pw} code~\cite{thermopw,Corso2016}.
The phonon frequencies were computed using the same LDA and PBE pseudopotentials as in the manuscript, without spin-orbit coupling, at nine different volumes. The resulting energies were fitted using the Murnaghan equation of state~\cite{Murnaghan1944}. 
We used a 18$\times$18$\times$18 $\mathbf{k}$-point grid for the electron and a 6$\times$6$\times$6 $\mathbf{q}$-point grid for the phonons. 
The obtained volume variation is given in Fig.~\ref{fig7}.

The change of eigenenergies due to thermal expansion is given by~\cite{Lautenschlager1985}:
\begin{equation}\label{renormeigen}
\Delta \varepsilon_{n\mathbf{k}} (T) = - \frac{\partial \varepsilon_{n\mathbf{k}}}{\partial P} \Big|_{T} \int_0^T dT' 3\alpha(T') B(T'),
\end{equation}
where $B(T) = -V(\partial P/\partial V)_T$ is the bulk modulus and $3\alpha = V^{-1}(\partial V/\partial T)_P$ is the 
thermal expansion coefficient. $B$ and $\alpha$ are obtained via numerical differentiation
starting from the volumes calculated in the above figure. 

We note that in Eq.~\ref{renormeigen} we carried the $\partial \varepsilon_{n\mathbf{k}}/\partial P$ term out of the temperature 
integral. This common temperature-independent approximation is valid in the elastic regime. 
To make sure that this approximation is valid, we compute $\partial \varepsilon_{n\mathbf{k}}/\partial P$ at 4~K and 300~K
by numerical derivation around the equilibrium volume for that temperature. From Table~\ref{table3}
 we see that indeed the temperature dependence is negligible and we therefore use the value at 4~K in Eq.~\ref{renormeigen}.

\begin{table}
  \begin{tabular}{l r r r r }
  \toprule
$\partial \varepsilon_{n\mathbf{k}}/\partial P$ & \multicolumn{2}{c}{LDA} & \multicolumn{2}{c}{PBE} \\
eV/Mbar                                         &   4 K       & 300 K     & 4 K       & 300 K \\
  \hline
%VBM     & 11.2573250666   &  11.2623783012  & 11.841529621       & 11.8745669506 \\
%CBM     &  9.53512857839  &  9.54115603027  &  9.85438063944     &  9.88407562831 \\
%Ind Gap & -1.72219648821  & -1.72122227093 & -1.98714898156 & -1.99049132229\\ 
VBM     & 11.257  &  11.262  & 11.841  & 11.874 \\
CBM     &  9.535  &   9.541  &  9.854  &  9.884 \\
Ind. Gap & -1.722  &  -1.721  & -1.987  & -1.990 \\ 
%Ind. Gap  exp.~\cite{Chang1984} &\multicolumn{4}{c}{-1.41, -1.5, -1.6, -3.8 }  \\
  \botrule 
  \end{tabular}
  \caption{\label{table3}
   Variation of the eigenenergies with pressure at two temperatures using the PBE and LDA pseudopotentials. %Experimental references are from Ref.~\onlinecite{Chang1984}.
  }
\end{table}

\begin{figure}
  \centering
  \includegraphics[width=\columnwidth]{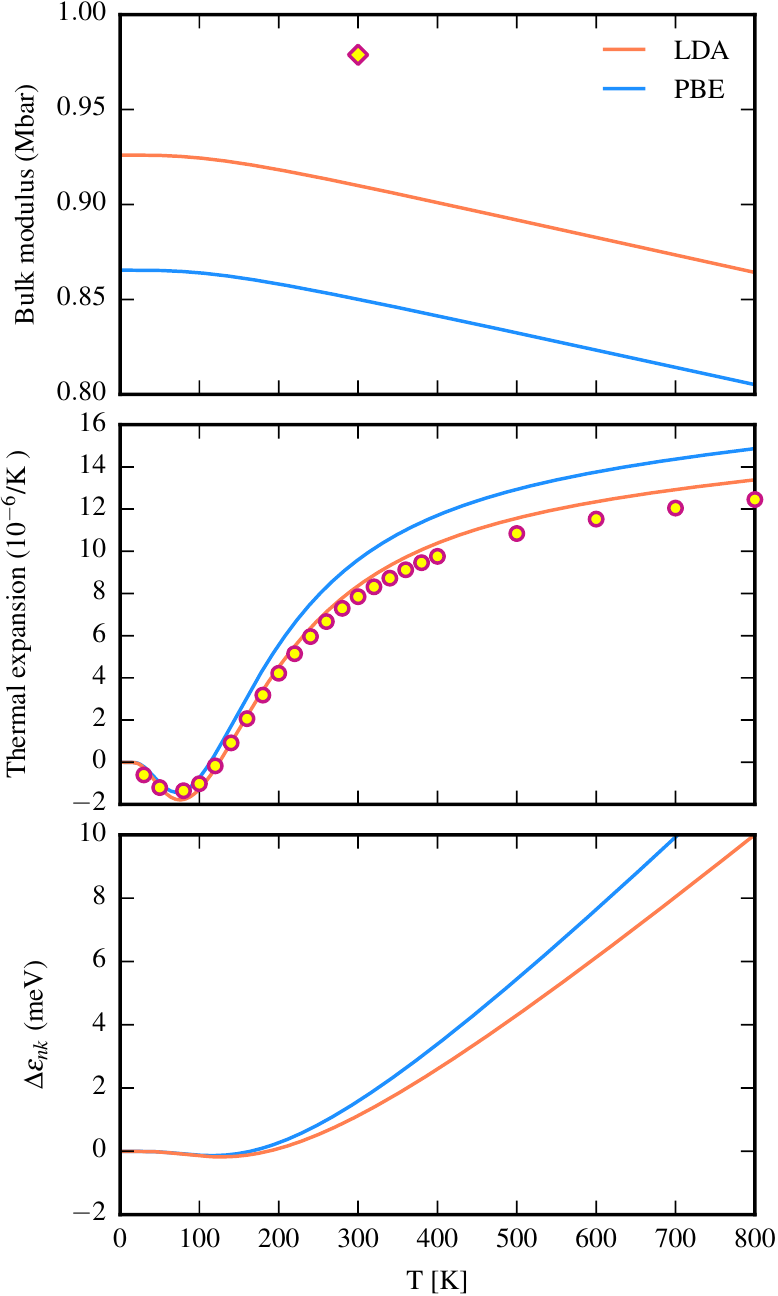}
  \caption{\label{fig8}
   Variation of the bulk modulus, the thermal expansion and eigenstates renormalization with temperature of silicon.
   Experimental values are from Refs.~\onlinecite{McSkimin1964a} (yellow diamond) and \onlinecite{Okada1984} (yellow dots).
   %Prior LDA calculation is also reported from Ref.~\onlinecite{Rignanese1995} (yellow square).
  }
\end{figure}

The bulk modulus, the thermal expansion and eigenstates renormalization with temperature of silicon computed with the LDA and PBE exchange-correlation 
functionals are presented in Fig.~\ref{fig8}. 
From the bottom panel we see that thermal lattice expansion leads to a slight increase of the band gap 
of silicon. 
For PBE, the valence band top and conduction band bottom change by $-$9.5~meV and 
$-$7.8~meV from 0~K to 300~K, therefore the net increase of the band gap is 1.7~meV. This variation
is much smaller than the gap renormalization arising from electron-phonon interactions (as discussed
next), therefore in the present case this effect can safely be neglected when calculating
carrier mobilities.

\subsection{Electron-phonon renormalization of the bandstructures and free carrier screening}

The electron-phonon renormalization of the bandstructure has been discussed 
for the case of silicon in considerable detail in Ref.~\onlinecite{Ponce2015}.
The calculated zero-point renormalization of the
fundamental gap is $-$56.2~meV within the non-adiabatic Rayleigh-Schr\"odinger perturbation theory. This
change corresponds to 5\% of the band gap. We extracted the effective masses using data 
from that paper, and the 
results are shown in Table~\ref{table4} for specific directions.

\begin{table}
  \begin{tabular}{c c c c c }
  \toprule
  Band & Direction       & without e-ph & with e-ph & Variation  \\
       &                 & interaction & interaction &   \\
  \hline
  \multirow{3}{*}{Light hole}
  & [100]                & 0.334 & 0.342 & +2\% \\
  & [111]                & 0.656 & 0.697 & +6\% \\
  \multirow{3}{*}{Heavy hole}
  & [100]                & 0.334 & 0.343 & +3\% \\
  & [111]                & 0.656 & 0.704 & +3\% \\
  \multirow{3}{*}{Split-off hole}
  & [100]                & 0.218 & 0.220 & +1\%  \\
  & [111]                & 0.099 & 0.100 & +1\%   \\
  Electron               & long.  & 0.927 & 0.966 & +4\% \\ 
  \botrule 
  \end{tabular}
  \caption{\label{table4}
  Effective masses of silicon computed at the LDA level without SOC, with or without
  the electron-phonon renormalization of the band structure at 0~K. The effective masses
  were obtained by using data from Ref.~\onlinecite{Ponce2015}.  
  }
\end{table}

The electron-phonon renormalization of the bands leads to an increase of both electron and hole effective 
masses between 1\% and 6\%. 
Therefore, we can reasonably estimate that similar changes in effective masses will occurs in our calculations.
Given the dependence of the mobility on effective mass, we estimate a 5\% reduction in mobility due to this effect. 

The effect of free carrier screening can be included via a
Lindhard dielectric function using {\it ab initio} parameters as described in
Ref.~\onlinecite{Verdi2017}. 
In this case the screening only affects phonons with energy below the
plasma energy of the doped carriers. %, since the screening is to be evaluated at each phonon frequency. 
For intrinsic silicon, which is the main focus of our work, the carrier concentration is below 
10$^{15}$~cm$^{-3}$. Using data from Ref.~\onlinecite{Caruso2016}, we estimate 
that the plasma energy in this case would be well below 0.1~meV, therefore the free carrier screening 
would be ineffective for nearly all phonons. We also mention that the renormalization of the band structure
arising from the free carriers is negligible in this case~\cite{Caruso2016}.

\bibliography{bibliography}

\end{document}